\documentclass[10pt,aps,pre,twocolumn,superscriptaddress,floatfix,longbibliography]{revtex4-2}

\usepackage{amsmath,amssymb,amsfonts,mathtools}
\usepackage{graphicx}
\usepackage{multirow}
\usepackage{appendix}
\usepackage{xcolor}
\usepackage{siunitx}
\usepackage{yfonts}
\usepackage{esint}
\usepackage{placeins}

\usepackage{hyperref}

\graphicspath{{./IMG_03/}}

\begin{document}

\title{A Fixed-Grid Affine-Constrained Multiwavelet Coefficient Method for Buckley--Leverett Shock Capturing}

\author{Christian Tantardini}
\email{christiantantardini@ymail.com}
\affiliation{Center for Integrative Petroleum Research, King Fahd University of Petroleum and Minerals, Dhahran 31261, Saudi Arabia.}

\author{Evgueni Dinvay}
\email{evgueni.dinvay@gmail.com}
\affiliation{Hylleraas Center, Department of Chemistry, UiT The Arctic University of Norway, PO Box 6050 Langnes, N-9037 Troms\o, Norway}

\date{\today}

\begin{abstract}
We present a fixed-grid conservative affine-constrained modal/multiwavelet
coefficient method for one-dimensional Buckley--Leverett saturation transport.
The saturation is evolved directly in a local orthonormal coefficient basis with
a mean/detail structure: the first mode carries the conservative cell average,
whereas higher modes carry zero-mean local details. The hyperbolic inflow
condition is imposed as a linear trace constraint on the coefficient vector and
enforced by affine lifting. For \(p>1\), the boundary reprojection is applied in
the detail subspace of the inflow cell, so that the prescribed trace is restored
without modifying the conservative cell-average update. The transport operator
is discretized in conservative weak form with monotone numerical fluxes, and
shock-induced oscillations are controlled by a troubled-cell limiter acting on
modal details.

The method is validated on a Berea-core waterflood benchmark against an
independent \texttt{pywaterflood} reference solution using the same Corey
fractional-flow closure, physical parameters, and pore-volume-injected scaling.
The affine-constrained coefficient solver reproduces the reference breakthrough
curve and saturation profiles, preserves the imposed inflow trace to roundoff
accuracy, controls saturation bounds through mean-preserving detail rescaling,
and gives small accumulated global mass-balance defects. Mesh-refinement,
flux-comparison, and modal-order studies show that \(p=2\), corresponding to a
piecewise-linear local representation, provides the most favorable
accuracy--cost compromise among the tested orders for this shock-dominated
benchmark.
\end{abstract}

\maketitle

\section{Introduction}
\label{sec:intro}

Accurate prediction of immiscible two-phase displacement is central to
reservoir engineering, because decisions on well placement, injection strategy,
breakthrough control, and sweep efficiency depend on reliable
saturation-front dynamics in porous media
\cite{BuckleyLeverett1942,BuckleyLeverett1952,belazreg2019novel}. At the
continuum level, these processes are commonly described by incompressible
two-phase flow, combining Darcy's law, phase mass conservation, and a
Buckley--Leverett-type saturation equation
\cite{Dake1978,Helmig1997,ChenHuanMa2006}. In the purely convective limit, the
saturation equation is hyperbolic and may generate entropy shocks and
rarefactions even from smooth initial data \cite{kaasschieter1999buckley}. When
capillary or dynamic-capillary effects are retained, the same transport
structure is regularized by parabolic or pseudo-parabolic terms
\cite{spayd2011buckley,vanDuijn2015dynamic,jin2016degenerate}.

This distinction has direct numerical consequences. In the one-dimensional
hyperbolic case with positive velocity, only the inflow boundary is prescribed,
whereas the outflow trace is determined by the interior solution. By contrast,
capillarity-regularized variants require additional boundary information. A
coefficient-space discretization must therefore distinguish the physical
imposition of boundary data from auxiliary analytical or numerical devices used
to construct operators. This point is especially important for wavelet,
multiwavelet, and modal formulations, where basis coefficients, traces,
projections, and reconstructed fields do not automatically satisfy pointwise
boundary constraints unless the admissible coefficient space is explicitly
enforced.

High-resolution saturation solvers are traditionally based on finite-volume
methods, WENO-type reconstructions, discontinuous Galerkin schemes, and adaptive
multiresolution strategies
\cite{Shu2016high,WENO_methods,Gerhard2015adaptiveMW,BMWDRS2020}. Wavelet and
multiwavelet techniques have also been used for adaptivity, compression,
stochastic representations, and uncertainty quantification
\cite{maitre_2017,Fouladi_2013,SEMBIRING2002211}. In adaptive multiresolution
schemes for conservation laws, the solution is represented on nested grid
hierarchies, and detail or prediction coefficients are used to activate,
discard, or refine resolution levels
\cite{Harten1994AdaptiveMR,HovhannisyanMullerSchafer2014,CaviedesVoullieme2020MWDG}.
Recent multiresolution discontinuous Galerkin approaches similarly use
hierarchical or multiwavelet representations to encode information across mesh
levels and guide adaptive refinement
\cite{Gerhard2022WaveletFree,HuangCheng2020MRDG}.

The present work develops a fixed-grid conservative coefficient-space solver
for Buckley--Leverett shock propagation. The saturation is represented by
cell-local orthonormal polynomial modes organized as a mean/detail hierarchy:
the first mode carries the conservative cell average, whereas the higher modes
carry zero-mean local detail information. This hierarchy is used as part of the
numerical formulation itself. The inflow boundary condition is imposed as an
affine constraint on the coefficients, oscillatory detail content is limited
near shocks, and the mean coefficients retain a finite-volume-like conservative
flux balance. All computations in this work are performed on fixed uniform
partitions; no dynamic mesh refinement, coefficient thresholding, or sparse
adaptive operator representation is used.

Although the weak residual is closely related to modal discontinuous Galerkin
discretizations, the formulation is organized around coefficient-space
operations rather than around reconstruction alone. The conservative mean mode
enters the telescoping flux balance, the zero-mean detail modes provide the
subcell degrees of freedom that are limited near shocks, and the inflow trace is
handled as a linear functional of the coefficient vector. This organization
keeps the boundary condition, limiter, and conservative update in the same set
of unknowns.

The main contribution is the affine treatment of the hyperbolic inflow
condition. The prescribed trace is written as a linear functional of the local
coefficient vector and imposed as an affine constraint on the admissible
coefficient space. For \(p>1\), where \(p\) is the number of local modes per
cell, the stage reprojection is applied in the detail subspace of the inflow
cell. This restores the imposed boundary trace without changing the mean
coefficient of the inflow cell and therefore without modifying the conservative
cell-average update.

The second ingredient is a conservative weak-form discretization of the
Buckley--Leverett flux. Neighboring cells are coupled by monotone numerical
interface fluxes, so that internal fluxes cancel pairwise in the global mass
balance. The third ingredient is a troubled-cell limiter acting on zero-mean
detail coefficients. The limiter suppresses nonphysical subcell oscillations
near Buckley--Leverett shocks while preserving the cell averages that carry the
conservative flux-difference update. The resulting method combines a
finite-volume-like conservation structure with a fixed-grid modal/multiwavelet
coefficient representation.

The method is validated on a Berea-core waterflood benchmark. The
affine-constrained coefficient solver is compared with an independent
\texttt{pywaterflood} reference calculation using the same Corey
fractional-flow closure, physical parameters, and pore-volume-injected scaling
\cite{Male2024pywaterflood}. The comparison includes breakthrough behavior,
saturation profiles, flux sensitivity, mesh-refinement behavior, modal-order
sensitivity, trace preservation, mass-balance diagnostics, and runtime.

The paper is organized as follows. Section~\ref{sec:phys-model} derives the
Buckley--Leverett transport model and its one-dimensional hyperbolic reduction.
Section~\ref{sec:mw-conservative} introduces the fixed-grid affine-constrained
modal/multiwavelet coefficient formulation, including the weak residual,
numerical fluxes, coefficient-space boundary constraint, limiter, and time
integration. Section~\ref{sec:numerics} presents the Berea-core validation
against the \texttt{pywaterflood} reference solution.
Section~\ref{sec:conclusions} summarizes the results and outlines extensions.

\section{Buckley--Leverett model}
\label{sec:phys-model}

\subsection{Two-phase flow equations}
\label{subsec:two-phase}

We consider immiscible, incompressible two-phase flow of a wetting phase \(w\)
and a non-wetting phase \(o\) in a rigid porous medium. The porosity is
\(\phi(\boldsymbol x)\), the absolute permeability tensor is
\(\boldsymbol K(\boldsymbol x)\), and the phase saturations satisfy
\begin{align}
    S_w=S,\qquad S_o=1-S .
\end{align}
The phase densities \(\rho_\ell\) and viscosities \(\mu_\ell\), with
\(\ell\in\{w,o\}\), are constant. The phase pressures are \(p_w\) and \(p_o\),
and the capillary pressure is defined by
\begin{align}
    p_c(S)=p_o-p_w .
\end{align}

Mass conservation for the two incompressible phases gives
\begin{align}
\phi\,\partial_t S + \nabla\!\cdot\boldsymbol u_w &= q_w,
\label{eq:water-balance}\\
\phi\,\partial_t(1-S) + \nabla\!\cdot\boldsymbol u_o &= q_o .
\label{eq:oil-balance}
\end{align}
The phase fluxes satisfy Darcy's law,
\begin{align}
\boldsymbol u_\ell
=
-\boldsymbol K\,\lambda_\ell(S)
\left(\nabla p_\ell-\rho_\ell\boldsymbol g\right),
\qquad
\ell\in\{w,o\},
\label{eq:darcy}
\end{align}
where
\begin{align}
    \lambda_\ell(S)=\frac{k_{r\ell}(S)}{\mu_\ell}
\end{align}
is the phase mobility.

We define the total velocity, total source, total mobility, and wetting-phase
fractional-flow function as
\begin{align}
    \boldsymbol v &= \boldsymbol u_w+\boldsymbol u_o,\\
    q_t &= q_w+q_o,\\
    \lambda_t(S) &= \lambda_w(S)+\lambda_o(S),\\
    f(S) &= \frac{\lambda_w(S)}{\lambda_t(S)} .
\end{align}
Adding Eqs.~\eqref{eq:water-balance} and \eqref{eq:oil-balance} gives the total
continuity equation
\begin{align}
    \nabla\!\cdot\boldsymbol v=q_t .
    \label{eq:total-continuity}
\end{align}

A standard way to formulate incompressible two-phase flow is to introduce a
global pressure, which absorbs the capillary contribution under the usual
global-pressure compatibility assumptions
\cite{ChaventJaffre1986,ChenHuanMa2006}. In the simplified setting relevant to
the Buckley--Leverett limit used below, the total velocity can be written as
\begin{align}
    \boldsymbol v=-\boldsymbol K\,\lambda_t(S)\,\nabla p_t,
    \label{eq:total-velocity}
\end{align}
where \(p_t\) is the global pressure. The corresponding pressure equation is
\begin{align}
    -\nabla\!\cdot\!\left(\boldsymbol K\,\lambda_t(S)\,\nabla p_t\right)
    =
    q_t .
    \label{eq:pressure-equation}
\end{align}
This pressure equation is not solved in the numerical experiments below; it is
included only to place the one-dimensional transport model in the standard
two-phase-flow setting.

The wetting-phase flux can be expressed in fractional-flow form as
\begin{align}
\boldsymbol u_w
=
f(S)\boldsymbol v
&+
\boldsymbol K\frac{\lambda_w(S)\lambda_o(S)}{\lambda_t(S)}\nabla p_c(S)
\nonumber\\
&-
\boldsymbol K\frac{\lambda_w(S)\lambda_o(S)}{\lambda_t(S)}
(\rho_o-\rho_w)\boldsymbol g .
\label{eq:wetting-flux}
\end{align}
Substitution into Eq.~\eqref{eq:water-balance} yields
\begin{widetext}
\begin{align}
\phi\,\partial_t S
+
\nabla\!\cdot\!\left(f(S)\boldsymbol v+\boldsymbol F_g(S)\right)
=
\nabla\!\cdot\!\left(
\boldsymbol K\frac{\lambda_w(S)\lambda_o(S)}{\lambda_t(S)}
\nabla p_c(S)
\right)
+
q_w ,
\label{eq:general-transport}
\end{align}
\end{widetext}
with
\begin{align}
\boldsymbol F_g(S)
=
-\boldsymbol K\frac{\lambda_w(S)\lambda_o(S)}{\lambda_t(S)}
(\rho_o-\rho_w)\boldsymbol g .
\label{eq:gravity-flux}
\end{align}

The computations in this work are restricted to the isothermal,
incompressible, one-dimensional Buckley--Leverett limit. Thus capillarity,
gravity, and phase sources are neglected in the transport step. Under these
assumptions, Eq.~\eqref{eq:general-transport} reduces to the classical
conservation law
\begin{align}
    \phi\,\partial_t S+\nabla\!\cdot\!\left(f(S)\boldsymbol v\right)=0 .
    \label{eq:BL-multid}
\end{align}
No thermal energy equation is solved in this work.

\subsection{One-dimensional hyperbolic reduction}
\label{subsec:oned-reduction}

We consider a one-dimensional core
\begin{align}
    \Omega=(0,L),
\end{align}
with
\begin{align}
    S=S(x,t),\qquad \boldsymbol v=(v,0,0),
\end{align}
where \(v>0\) is constant. Equation~\eqref{eq:BL-multid} then becomes
\begin{align}
    \phi\,\partial_t S+\partial_x\!\left(vf(S)\right)=0 .
    \label{eq:BL-1D-cons}
\end{align}
Equivalently, after division by the constant porosity \(\phi\), the equation is
written as
\begin{align}
    \partial_t S+\partial_x\mathcal F(S)=0,
    \qquad
    \mathcal F(S)=\frac{v}{\phi}f(S).
    \label{eq:BL-scaled-flux}
\end{align}
This is the conservative form discretized by the multiwavelet weak formulation
below.

For the hyperbolic problem with \(v>0\), the physical boundary datum is
prescribed only at the inflow boundary:
\begin{align}
    S(0,t)=g_D(t).
    \label{eq:inflow-bc}
\end{align}
The point \(x=L\) is an outflow boundary. No Dirichlet condition is prescribed
there in the purely hyperbolic case, because the trace \(S(L,t)\) is determined
by the interior solution. Imposing an additional Dirichlet value at \(x=L\)
would overconstrain the first-order hyperbolic problem. If capillarity or
another genuinely parabolic regularization is retained, the boundary setting
must be enlarged accordingly.

We reserve \(f(S)\) for the fractional-flow function and denote the prescribed
boundary datum by \(g_D(t)\). In the Berea-core computations reported below,
the inflow datum is the injected-water saturation,
\begin{align}
    g_D(t)=S_{\rm inj}=1-S_{or},
    \label{eq:injected-saturation}
\end{align}
where \(S_{or}\) is the residual oil saturation.

\section{Fixed-grid affine-constrained multiwavelet coefficient formulation}
\label{sec:mw-conservative}

We discretize the one-dimensional conservation law \eqref{eq:BL-scaled-flux}
directly in a local orthonormal modal/multiwavelet coefficient space.
The formulation is
conservative in weak form, so that discontinuities are coupled through
numerical interface fluxes rather than through a strong derivative of the
reconstructed solution. Its organization is coefficient-space rather than
reconstruction-space: boundary traces, cell averages, detail modes, limiting,
and affine constraints are all expressed in terms of the local modal
coefficients.

The method has three structural features. First, the physical inflow datum is
imposed as an affine trace constraint on the coefficient vector. Second, the
local basis separates the conservative mean mode from zero-mean detail modes,
which provides a natural target for shock limiting. Third, for \(p>1\), the
boundary reprojection is applied only to detail coefficients in the inflow cell,
so that the imposed trace is recovered without changing the first-cell average.
These choices preserve the conservative flux-difference structure on the fixed
grid and define the coefficient-space operations validated in the numerical
experiments below.

\subsection{Local multiwavelet approximation}
\label{subsec:cellwise-mw}

Let \(\Omega=(0,L)\) be partitioned into \(N_c\) cells
\begin{align}
    I_c&=(x_{c-1/2},x_{c+1/2}), \\
    h_c&=x_{c+1/2}-x_{c-1/2}, \\
    c&=1,\ldots,N_c .
\end{align}
Each cell is mapped to the reference interval \(\widehat I=(-1,1)\) by
\begin{align}
    x=x_c+\frac{h_c}{2}\xi,
    \qquad
    x_c=\frac{x_{c-1/2}+x_{c+1/2}}{2},
    \label{eq:cell-map}
\end{align}
with \(\xi\in(-1,1)\). On \(\widehat I\), let \(P_k(\xi)\) denote the Legendre
polynomial of degree \(k\). The local cell-supported basis used in the
implementation is
\begin{align}
    \psi_{c,k}(x)
    =
    \left(\frac{2k+1}{h_c}\right)^{1/2}
    P_k\!\left(\frac{2(x-x_c)}{h_c}\right),
    \label{eq:local-basis}
\end{align}
with $k=0,\ldots,p-1$. Here, \(p\) denotes the number of local modes per cell; the highest polynomial
degree is therefore \(p-1\). The normalization in Eq.~\eqref{eq:local-basis}
gives the orthonormality relation
\begin{align}
    \int_{I_c}\psi_{c,k}(x)\psi_{c,\ell}(x)\,dx
    =
    \delta_{k\ell},
    \qquad
    k,\ell=0,\ldots,p-1 .
    \label{eq:cell-orthonormality}
\end{align}
Accordingly, the local mass matrix is the identity,
\begin{align}
    M_{k\ell}^{(c)}
    =
    \int_{I_c}\psi_{c,k}(x)\psi_{c,\ell}(x)\,dx
    =
    \delta_{k\ell},
    \label{eq:identity-mass}
\end{align}
so no local mass inversion is required after assembling the weak residual.

This cellwise orthonormal polynomial representation is used here as a local
modal/multiwavelet coefficient representation. The terminology is consistent
with piecewise-polynomial multiwavelet constructions, while the conservative
weak-form treatment follows the standard structure of high-order discontinuous
formulations for conservation laws
\cite{Alpert1993,CockburnShu2001RKDG,HesthavenWarburton2008}. The important
point for the present fixed-grid formulation is the mean/detail coefficient
structure: the first mode carries the conservative cell average, whereas the
higher modes carry zero-mean local detail information.

The numerical saturation is expanded as
\begin{align}
    S_h(x,t)
    =
    \sum_{c=1}^{N_c}\sum_{k=0}^{p-1}
    s_{c,k}(t)\psi_{c,k}(x),
    \qquad x\in I_c .
    \label{eq:cellwise-mw-expansion}
\end{align}
Equivalently, on a fixed cell,
\begin{align}
    S_h|_{I_c}(x,t)
    =
    \sum_{k=0}^{p-1}s_{c,k}(t)\psi_{c,k}(x).
    \label{eq:local-expansion}
\end{align}
The evolved degrees of freedom are therefore the local modal coefficients
\(s_{c,k}(t)\).

The first basis function is the mean mode. Since
\(\psi_{c,0}=h_c^{-1/2}\), the cell average is
\begin{align}
    \overline S_c(t)
    =
    \frac{1}{h_c}\int_{I_c}S_h(x,t)\,dx
    =
    \frac{s_{c,0}(t)}{\sqrt{h_c}} .
    \label{eq:mean-coeff-relation}
\end{align}
All higher modes have zero cell average,
\begin{align}
    \int_{I_c}\psi_{c,k}(x)\,dx=0,
    \qquad k=1,\ldots,p-1,
    \label{eq:zero-mean-modes}
\end{align}
and therefore represent local detail coefficients. Hence
\begin{align}
    S_h|_{I_c}(x,t)
    =
    \overline S_c(t)+S_c'(x,t),
    \label{eq:mean-detail-local}
\end{align}
where
\begin{align}
    \int_{I_c}S_c'(x,t)\,dx=0,
\end{align}
and
\begin{align}
    S_c'(x,t)
    =
    \sum_{k=1}^{p-1}s_{c,k}(t)\psi_{c,k}(x).
    \label{eq:detail-expansion}
\end{align}
This mean/detail split is central to the method. The mean coefficient carries
the finite-volume-like conservative quantity, while the higher coefficients
carry subcell information that can be limited selectively near shocks.
This mean/detail structure is also the natural coefficient organization for
future adaptive multiresolution extensions, although no adaptive refinement is
used in the computations reported here.

Because the basis is supported cellwise, \(S_h\) is generally discontinuous at
cell interfaces. The left and right traces on cell \(I_c\) are
\begin{align}
    S_{c-1/2}^{+}(t)
    &=
    \sum_{k=0}^{p-1}s_{c,k}(t)\psi_{c,k}(x_{c-1/2}^{+}),
    \label{eq:left-trace-local}\\
    S_{c+1/2}^{-}(t)
    &=
    \sum_{k=0}^{p-1}s_{c,k}(t)\psi_{c,k}(x_{c+1/2}^{-}).
    \label{eq:right-trace-local}
\end{align}
Using \(P_k(1)=1\) and \(P_k(-1)=(-1)^k\), the trace values of the basis are
\begin{align}
    \psi_{c,k}(x_{c+1/2}^{-})
    &=
    \left(\frac{2k+1}{h_c}\right)^{1/2},
    \label{eq:right-trace-basis}\\
    \psi_{c,k}(x_{c-1/2}^{+})
    &=
    (-1)^k
    \left(\frac{2k+1}{h_c}\right)^{1/2}.
    \label{eq:left-trace-basis}
\end{align}
These traces enter both the numerical interface fluxes and the
coefficient-space inflow constraint. In particular, at the left boundary
\(x=0=x_{1-1/2}\), the prescribed inflow condition becomes a linear constraint
on the coefficients of the first cell.

Volume integrals are evaluated by Gauss--Legendre quadrature. If
\(\{\xi_q,w_q\}_{q=1}^{Q}\) are the quadrature nodes and weights on
\((-1,1)\), then
\begin{align}
    x_{c,q}=x_c+\frac{h_c}{2}\xi_q,
    \qquad
    w_{c,q}=\frac{h_c}{2}w_q .
    \label{eq:cell-quadrature-map}
\end{align}
For a nonlinear function \(G(S_h)\), we use
\begin{align}
    \int_{I_c}G(S_h(x,t))\,\psi_{c,k}(x)\,dx
    \approx
    \sum_{q=1}^{Q}
    w_{c,q}
    G(S_h(x_{c,q},t))
    \psi_{c,k}(x_{c,q}).
    \label{eq:cell-quadrature}
\end{align}
The same quadrature nodes are used to evaluate the nonlinear fractional-flow
flux in the volume term and to monitor saturation bounds inside each cell.

In the implementation, the values of
\(\psi_{c,k}(x_{c,q})\), \(\partial_x\psi_{c,k}(x_{c,q})\), and the left and
right trace vectors are precomputed. Residual assembly therefore consists of
coefficient-to-node evaluation, pointwise flux evaluation, projection of the
volume term back to coefficient space, and conservative coupling through
interface fluxes. This local modal/multiwavelet coefficient structure enables
the combination of coefficient-space boundary enforcement, conservative flux
coupling, and detail-selective limiting used in the following subsections.

\subsection{Conservative weak residual}
\label{subsec:conservative-weak-residual}

The equation discretized in the computations is the scalar conservation law
\begin{align}
    \partial_t S+\partial_x\mathcal F(S)=0,
    \qquad
    \mathcal F(S)=\frac{v}{\phi}f(S),
    \label{eq:mw-transport-equation}
\end{align}
where \(f(S)=\lambda_w(S)/[\lambda_w(S)+\lambda_o(S)]\) is the wetting-phase
fractional-flow function. The discretization follows the weak conservative
structure used in high-order discontinuous formulations for hyperbolic
conservation laws~\cite{CockburnShu2001RKDG,CockburnShu1989RKDG}. In the
present formulation, however, the residual is assembled and evolved directly in
the local modal/multiwavelet coefficient space introduced above.

For each cell \(I_c\) and local basis function \(\psi_{c,k}\),
\(k=0,\ldots,p-1\), multiplication of
Eq.~\eqref{eq:mw-transport-equation} by \(\psi_{c,k}\), integration over
\(I_c\), and integration by parts give
\begin{widetext}
\begin{align}
\frac{d}{dt}
\int_{I_c} S_h(x,t)\psi_{c,k}(x)\,dx
=
\int_{I_c}\mathcal F(S_h(x,t))\,\partial_x\psi_{c,k}(x)\,dx
-
\widehat{\mathcal F}_{c+1/2}\psi_{c,k}(x_{c+1/2}^{-})
+
\widehat{\mathcal F}_{c-1/2}\psi_{c,k}(x_{c-1/2}^{+}) .
\label{eq:mw-weak-residual}
\end{align}
\end{widetext}
Here \(\widehat{\mathcal F}_{c\pm1/2}\) are numerical fluxes at cell
interfaces. They replace the generally double-valued physical flux
\(\mathcal F(S_h)\) at discontinuities of the cellwise representation and
provide the conservative coupling between neighboring coefficient blocks.

Using the orthonormality relation \eqref{eq:cell-orthonormality}, the left-hand
side of Eq.~\eqref{eq:mw-weak-residual} is simply \(d s_{c,k}/dt\). Therefore
the coefficient residual is
\begin{align}
    \frac{d s_{c,k}}{dt}
    =
    R_{c,k}(\boldsymbol s),
    \label{eq:coefficient-ode}
\end{align}
with
\begin{align}
R_{c,k}(\boldsymbol s)
=
V_{c,k}(\boldsymbol s)
&-
\widehat{\mathcal F}_{c+1/2}\psi_{c,k}(x_{c+1/2}^{-}) \nonumber \\
&+
\widehat{\mathcal F}_{c-1/2}\psi_{c,k}(x_{c-1/2}^{+}),
\label{eq:coefficient-residual}
\end{align}
where the volume contribution is
\begin{align}
    V_{c,k}(\boldsymbol s)
    =
    \int_{I_c}
    \mathcal F(S_h(x,t))\,\partial_x\psi_{c,k}(x)\,dx .
    \label{eq:volume-residual}
\end{align}
In the implementation, this integral is evaluated by the same Gauss--Legendre
quadrature used in Eq.~\eqref{eq:cell-quadrature}:
\begin{align}
    V_{c,k}(\boldsymbol s)
    \approx
    \sum_{q=1}^{Q}
    w_{c,q}\,
    \mathcal F(S_h(x_{c,q},t))\,
    \partial_x\psi_{c,k}(x_{c,q}).
    \label{eq:volume-residual-quadrature}
\end{align}
Thus residual assembly consists of evaluating \(S_h\) at quadrature points,
computing the nonlinear flux pointwise, projecting the volume term against
\(\partial_x\psi_{c,k}\), and adding the conservative interface flux
contributions.

At an internal interface \(x_{c+1/2}\), the numerical flux depends on the left
and right traces
\begin{align}
    S_{c+1/2}^{-}
    &=
    S_h(x_{c+1/2}^{-},t),\\
    S_{c+1/2}^{+}
    &=
    S_h(x_{c+1/2}^{+},t).
\end{align}
We write
\begin{align}
    \widehat{\mathcal F}_{c+1/2}
    =
    \widehat{\mathcal F}
    \left(
    S_{c+1/2}^{-},
    S_{c+1/2}^{+}
    \right),
    \label{eq:numerical-flux}
\end{align}
where \(\widehat{\mathcal F}\) is consistent,
\begin{align}
    \widehat{\mathcal F}(S,S)=\mathcal F(S).
    \label{eq:flux-consistency}
\end{align}

In the production Berea-core calculations reported in
Sec.~\ref{sec:numerics}, we use the Rusanov flux~\cite{Rusanov1961},
\begin{align}
\widehat{\mathcal F}^{\rm Rus}_{c+1/2}
&=
\frac{1}{2}
\left[
\mathcal F(S_{c+1/2}^{-})
+
\mathcal F(S_{c+1/2}^{+})
\right]
\nonumber\\
&\quad
-
\frac{\alpha_{c+1/2}}{2}
\left(
S_{c+1/2}^{+}
-
S_{c+1/2}^{-}
\right),
\label{eq:rusanov-flux}
\end{align}
where
\begin{align}
    \alpha_{c+1/2}
    \geq
    \max_{S\in I(S_{c+1/2}^{-},S_{c+1/2}^{+})}
    |\mathcal F'(S)|.
    \label{eq:local-speed-bound}
\end{align}
Here \(I(S_L,S_R)\) denotes the closed interval with endpoints \(S_L\) and
\(S_R\). In the numerical experiments below, we use the global admissible-speed
bound
\begin{align}
    \alpha_{c+1/2}=a_{\max}
    =
    \max_{S\in[S_{wc},1-S_{or}]}
    |\mathcal F'(S)|,
    \label{eq:global-rusanov-speed}
\end{align}
which provides a robust dissipative baseline over the physical saturation
interval. The Rusanov flux is commonly used as a local Lax--Friedrichs-type
flux in finite-volume and discontinuous Galerkin discretizations of hyperbolic
problems. 

The implementation also permits a sampled Godunov flux~\cite{Godunov1959}. For
a scalar flux, this can be written as
\begin{align}
\widehat{\mathcal F}^{\rm God}(S_L,S_R)
=
\begin{cases}
\displaystyle
\min_{S\in[S_L,S_R]}\mathcal F(S),
&
S_L\leq S_R,\\[0.8em]
\displaystyle
\max_{S\in[S_R,S_L]}\mathcal F(S),
&
S_L>S_R.
\end{cases}
\label{eq:godunov-flux}
\end{align}
In the code, the extrema are evaluated by sampling the admissible saturation
interval. The sampled Godunov flux is used in
Sec.~\ref{subsec:flux-modal-resolution} to quantify the accuracy--cost
tradeoff relative to the Rusanov baseline.

The boundary fluxes are treated consistently with the hyperbolic character of
the problem. For \(v>0\), the left boundary is an inflow boundary and uses the
prescribed injected saturation,
\begin{align}
    \widehat{\mathcal F}_{1/2}
    =
    \widehat{\mathcal F}
    \left(
    S_{\rm inj},
    S_{1/2}^{+}
    \right).
    \label{eq:left-boundary-flux}
\end{align}
The right boundary is an outflow boundary; no Dirichlet value is imposed there.
The corresponding numerical flux is evaluated from the interior trace,
\begin{align}
    \widehat{\mathcal F}_{N_c+1/2}
    =
    \mathcal F(S_{N_c+1/2}^{-}),
    \label{eq:right-boundary-flux}
\end{align}
or equivalently by a consistent numerical flux with identical left and right
states.

The conservative structure is seen by choosing the mean mode \(k=0\). Since
\(\partial_x\psi_{c,0}=0\) and \(\psi_{c,0}=h_c^{-1/2}\),
Eq.~\eqref{eq:mw-weak-residual} gives
\begin{align}
    \frac{d\overline S_c}{dt}
    =
    -\frac{1}{h_c}
    \left(
    \widehat{\mathcal F}_{c+1/2}
    -
    \widehat{\mathcal F}_{c-1/2}
    \right).
    \label{eq:cell-average-update}
\end{align}
Thus the mean coefficients obey an exactly conservative flux-difference update,
while the higher detail modes evolve through the same weak residual. Summing
Eq.~\eqref{eq:cell-average-update} over all cells gives the global balance
\begin{align}
    \frac{d}{dt}
    \sum_{c=1}^{N_c} h_c\overline S_c
    =
    -
    \widehat{\mathcal F}_{N_c+1/2}
    +
    \widehat{\mathcal F}_{1/2}.
    \label{eq:global-conservation}
\end{align}
All internal numerical fluxes cancel pairwise. Therefore conservation is lost
neither through the local modal/multiwavelet representation nor through the
discontinuity of \(S_h\) at interfaces; the only contributions to the global
mass balance are the physical inflow and outflow fluxes.

\subsection{Affine enforcement of the inflow condition}
\label{subsec:boundary-lifting}

For \(v>0\), the one-dimensional Buckley--Leverett problem has a single
physical Dirichlet datum, namely the inflow trace
\begin{align}
    S(0,t)=g_D(t).
    \label{eq:physical-inflow-trace}
\end{align}
No Dirichlet value is imposed at \(x=L\), because the right boundary is an
outflow boundary and its trace is determined by the interior solution. In a
standard discontinuous weak formulation, boundary data enter through traces and
numerical fluxes \cite{CockburnShu2001RKDG,HesthavenWarburton2008}. Here, in
addition, the prescribed inflow trace is enforced algebraically as a linear
constraint on the local modal/multiwavelet coefficient vector. Thus the
admissible numerical states form an affine subspace of coefficient space.

Let the global coefficient vector be ordered as
\begin{widetext}
\begin{align}
    \boldsymbol s(t)
    =
    \left(
    s_{1,0},\ldots,s_{1,p-1},
    s_{2,0},\ldots,s_{2,p-1},
    \ldots,
    s_{N_c,0},\ldots,s_{N_c,p-1}
    \right)^{\mathsf T}.
    \label{eq:global-coeff-vector}
\end{align}
\end{widetext}
The left trace is a linear functional of \(\boldsymbol s\):
\begin{align}
    S_h(0^+,t)
    =
    \boldsymbol m^{\mathsf T}\boldsymbol s(t).
    \label{eq:left-trace-functional}
\end{align}
Only the entries associated with the first cell are nonzero. Using
Eq.~\eqref{eq:left-trace-basis}, one has
\begin{align}
    m_{1,k}
    =
    \psi_{1,k}(x_{1/2}^{+})
    =
    (-1)^k
    \left(\frac{2k+1}{h_1}\right)^{1/2},
    \label{eq:left-trace-vector}
\end{align}
with $k=0,\ldots,p-1$, and \(m_{c,k}=0\) for \(c>1\). Hence the inflow condition is
\begin{align}
    \boldsymbol m^{\mathsf T}\boldsymbol s(t)=g_D(t).
    \label{eq:trace-constraint-single}
\end{align}
Equivalently, with the row matrix \(\boldsymbol M=\boldsymbol m^{\mathsf T}\),
\begin{align}
    \boldsymbol M\boldsymbol s(t)=\boldsymbol g(t),
    \label{eq:trace-constraint-general}
\end{align}
where, for the present scalar inflow condition, \(\boldsymbol g(t)=g_D(t)\).

The admissible coefficient vectors form the affine space
\begin{align}
    \mathcal A_g(t)
    =
    \left\{
    \boldsymbol s:\boldsymbol M\boldsymbol s=\boldsymbol g(t)
    \right\}.
    \label{eq:affine-space}
\end{align}
A standard constrained parametrization is obtained by choosing a lifting vector
\(\boldsymbol\ell(t)\) such that
\begin{align}
    \boldsymbol M\boldsymbol\ell(t)=\boldsymbol g(t),
    \label{eq:lifting-vector}
\end{align}
and writing
\begin{align}
    \boldsymbol s(t)=\boldsymbol\ell(t)+\boldsymbol z(t),
    \qquad
    \boldsymbol M\boldsymbol z(t)=0 .
    \label{eq:affine-decomposition}
\end{align}
If the columns of \(\boldsymbol Z\) span \(\ker\boldsymbol M\), then
\begin{align}
    \boldsymbol z(t)=\boldsymbol Z\boldsymbol y(t),
    \qquad
    \boldsymbol s(t)=\boldsymbol\ell(t)+\boldsymbol Z\boldsymbol y(t),
    \label{eq:nullspace-param}
\end{align}
where \(\boldsymbol y(t)\) is an unconstrained reduced coefficient vector.

Let the conservative residual generated by Eq.~\eqref{eq:mw-weak-residual} be
\begin{align}
    \frac{d\boldsymbol s}{dt}
    =
    \boldsymbol{\mathcal R}(\boldsymbol s).
    \label{eq:generic-semidiscrete}
\end{align}
Substitution of Eq.~\eqref{eq:nullspace-param} gives
\begin{align}
    \boldsymbol Z\frac{d\boldsymbol y}{dt}
    =
    \boldsymbol{\mathcal R}
    \!\left(\boldsymbol\ell+\boldsymbol Z\boldsymbol y\right)
    -
    \frac{d\boldsymbol\ell}{dt}.
    \label{eq:lifted-preprojection}
\end{align}
Projection onto the homogeneous trace space gives
\begin{align}
    \frac{d\boldsymbol y}{dt}
    =
    \boldsymbol Z^\dagger
    \left[
    \boldsymbol{\mathcal R}
    \!\left(\boldsymbol\ell+\boldsymbol Z\boldsymbol y\right)
    -
    \frac{d\boldsymbol\ell}{dt}
    \right],
    \label{eq:lifted-evolution}
\end{align}
where \(\boldsymbol Z^\dagger\) is the left inverse associated with the chosen
coefficient inner product. If the columns of \(\boldsymbol Z\) are orthonormal
in this inner product, then \(\boldsymbol Z^\dagger=\boldsymbol Z^{\mathsf T}\).

In the implementation, \(\boldsymbol Z\) is not formed explicitly. Since there
is only one scalar inflow constraint, an equivalent correction can be applied
directly to the coefficient vector. For an arbitrary coefficient vector
\(\boldsymbol s\), define the trace defect
\begin{align}
    d_g(\boldsymbol s)
    =
    g_D(t)-\boldsymbol m^{\mathsf T}\boldsymbol s .
    \label{eq:trace-defect}
\end{align}
The full minimum-norm affine correction is
\begin{align}
    \mathcal P_{\mathcal A_g}^{\rm full}\boldsymbol s
    =
    \boldsymbol s
    +
    \frac{d_g(\boldsymbol s)}
    {\boldsymbol m^{\mathsf T}\boldsymbol m}
    \boldsymbol m .
    \label{eq:min-norm-affine-correction}
\end{align}
Indeed,
\begin{align}
    \boldsymbol m^{\mathsf T}
    \mathcal P_{\mathcal A_g}^{\rm full}\boldsymbol s
    =
    \boldsymbol m^{\mathsf T}\boldsymbol s
    +
    d_g(\boldsymbol s)
    =
    g_D(t),
    \label{eq:affine-correction-proof}
\end{align}
so \(\mathcal P_{\mathcal A_g}^{\rm full}\boldsymbol s\in\mathcal A_g(t)\).

A full correction enforces the trace, but it may also change the mean
coefficient of the inflow cell. For the production calculations with \(p>1\),
we therefore use a detail-only affine correction. Let \(\boldsymbol m_d\)
collect the trace values of the nonzero detail modes \(k=1,\ldots,p-1\) in the
first cell, and let \(\boldsymbol s_{1,d}\) denote the corresponding detail
coefficients. The detail-only correction is
\begin{align}
\boldsymbol s_{1,d}^{\,\rm new}
=
\boldsymbol s_{1,d}^{\,\rm old}
+
\frac{
g_D(t)-\boldsymbol m^{\mathsf T}\boldsymbol s^{\rm old}
}{
\boldsymbol m_d^{\mathsf T}\boldsymbol m_d
}
\boldsymbol m_d .
\label{eq:detail-affine-correction}
\end{align}
The mean coefficient \(s_{1,0}\) is left unchanged. Thus, for \(p>1\), the
affine reprojection restores the inflow trace without modifying the
conservative cell-average update in the inflow cell. For \(p=1\), no detail
subspace is available, and the implementation falls back to the full correction
in Eq.~\eqref{eq:min-norm-affine-correction}.

The same projection idea is applied to the residual. To keep the semidiscrete
velocity tangent to the homogeneous constraint, we remove the component of
\(\boldsymbol{\mathcal R}\) in the full trace direction:
\begin{align}
    \boldsymbol{\mathcal R}_{\rm tan}(\boldsymbol s)
    =
    \boldsymbol{\mathcal R}(\boldsymbol s)
    -
    \frac{
    \boldsymbol m^{\mathsf T}\boldsymbol{\mathcal R}(\boldsymbol s)
    }{
    \boldsymbol m^{\mathsf T}\boldsymbol m
    }
    \boldsymbol m .
    \label{eq:tangent-residual-single}
\end{align}
Consequently,
\begin{align}
    \boldsymbol m^{\mathsf T}
    \boldsymbol{\mathcal R}_{\rm tan}(\boldsymbol s)=0.
    \label{eq:tangent-residual-single-proof}
\end{align}
Thus the residual update is tangent to the homogeneous constraint manifold.
The affine correction is nevertheless reimposed after each Runge--Kutta stage
to remove roundoff-level drift and to ensure that the limited state remains
compatible with the prescribed trace.
For \(p>1\), this stage reprojection uses the detail-only correction in
Eq.~\eqref{eq:detail-affine-correction}. This choice preserves the mean
coefficient of the inflow cell and therefore does not alter the conservative
cell-average update. It should be distinguished from a pointwise bound
projection: the detail-only trace correction enforces the boundary trace under a
fixed cell average, and therefore pointwise boundedness of all monitored points
inside the constrained inflow cell is not guaranteed after the final
reprojection.

For the Berea-core computations,
\begin{align}
    g_D(t)=S_{\rm inj}=1-S_{or}
\end{align}
is constant. Hence \(d\boldsymbol\ell/dt=0\), and the role of the affine
lifting is to keep the left trace equal to the injected-water saturation
throughout the conservative coefficient evolution. The right boundary is not
included in \(\boldsymbol M\); it is treated as an outflow boundary through the
numerical flux in Eq.~\eqref{eq:right-boundary-flux}.

\subsection{Limiter and bound preservation}
\label{subsec:limiting}

The conservative residual \eqref{eq:mw-weak-residual} gives the correct
cell-average flux balance, but higher modal/detail coefficients must be
controlled near entropy shocks. We therefore apply a local limiting procedure
after each stage of the strong-stability-preserving Runge--Kutta update. The
construction follows the standard logic of local projection limiting for
discontinuous Galerkin conservation-law discretizations and
maximum-principle-preserving rescaling for scalar conservation
laws~\cite{CockburnShu1989RKDG,ZhangShu2010MaximumPrinciple,GottliebShuTadmor2001}.

Using the mean/detail decomposition \eqref{eq:mean-detail-local}, the limiter
acts on the zero-mean component \(S_c'\). Hence the cell average
\(\overline S_c\), and therefore the conservative mean update generated by
Eq.~\eqref{eq:mw-weak-residual}, is not changed by the detail limiting step.
This is the coefficient-space analogue of preserving finite-volume cell
averages while suppressing only the oscillatory subcell content responsible for
nonphysical oscillations near shocks.

Troubled cells are identified from the reconstructed interface traces. Let
\begin{align}
    S_c^- = S_h(x_{c-1/2}^{+},t),
    \qquad
    S_c^+ = S_h(x_{c+1/2}^{-},t),
\end{align}
and define the neighboring average jumps
\begin{align}
    \Delta_c^-=\overline S_c-\overline S_{c-1},
    \qquad
    \Delta_c^+=\overline S_{c+1}-\overline S_c .
\end{align}
A cell is flagged when the deviation of either trace from the cell average is
not compatible with the limited one-sided differences. Specifically, we compare
\begin{align}
    S_c^+ - \overline S_c
    \quad \text{with} \quad
    \operatorname{minmod}
    \left(
    S_c^+ - \overline S_c,\,
    \beta\Delta_c^-,\,
    \beta\Delta_c^+
    \right),
    \label{eq:troubled-right}
\end{align}
and
\begin{align}
    \overline S_c - S_c^-
    \quad \text{with} \quad
    \operatorname{minmod}
    \left(
    \overline S_c - S_c^-,\,
    \beta\Delta_c^-,\,
    \beta\Delta_c^+
    \right).
    \label{eq:troubled-left}
\end{align}
Here \(1\leq\beta\leq2\) is a sensitivity parameter, and
\begin{widetext}
\begin{align}
\operatorname{minmod}(a_1,\ldots,a_m)
=
\begin{cases}
\displaystyle
\operatorname{sgn}(a_1)\min_{1\leq j\leq m}|a_j|,
&
\text{if } \operatorname{sgn}(a_1)=\cdots=\operatorname{sgn}(a_m),\\[0.4em]
0,
&
\text{otherwise}.
\end{cases}
\label{eq:minmod}
\end{align}
\end{widetext}

Smooth cells are left unchanged by the troubled-cell limiter. In troubled
cells, the zero-mean detail part is rescaled according to
\begin{align}
    S_h^{\rm lim}(x,t)|_{I_c}
    =
    \overline S_c(t)+\theta_c S_c'(x,t),
    \qquad
    0\leq\theta_c\leq1 .
    \label{eq:detail-limiter}
\end{align}
The rescaling parameter is then chosen to enforce the admissible saturation
interval at the quadrature and interface points used by the scheme.

The admissible bounds are
\begin{align}
    S_{\min}=S_{wc},
    \qquad
    S_{\max}=1-S_{or}.
\end{align}
Let
\begin{align}
    S_c^{\max}
    =
    \max_{x_q\in I_c}S_h(x_q,t),
    \qquad
    S_c^{\min}
    =
    \min_{x_q\in I_c}S_h(x_q,t),
    \label{eq:local-extrema}
\end{align}
where the set of points includes the quadrature nodes used for volume
integration and the two cell interfaces. The bound-preserving rescaling
parameter is chosen as
\begin{align}
\theta_c
=
\min\left\{
1,\,
\frac{S_{\max}-\overline S_c}{S_c^{\max}-\overline S_c+\varepsilon},\,
\frac{\overline S_c-S_{\min}}{\overline S_c-S_c^{\min}+\varepsilon}
\right\},
\label{eq:theta-bound}
\end{align}
with a small \(\varepsilon>0\) to avoid division by zero. If the reconstructed
polynomial already lies in the admissible interval at the monitored points,
then \(\theta_c=1\). If not, only the zero-mean detail part is reduced.

For every monitored point \(x_q\in I_c\), namely every quadrature or interface
point used by the scheme, the limited value is
\begin{align}
    S_h^{\rm lim}(x_q,t)
    =
    \overline S_c
    +
    \theta_c\left(S_h(x_q,t)-\overline S_c\right).
    \label{eq:limited-point-value}
\end{align}
Therefore, by construction,
\begin{align}
    S_{wc}
    \leq
    S_h^{\rm lim}(x_q,t)
    \leq
    1-S_{or}
    \qquad
    \text{at all monitored points}.
    \label{eq:bound-preserving}
\end{align}
At the same time,
\begin{align}
    \frac{1}{|I_c|}
    \int_{I_c}S_h^{\rm lim}(x,t)\,dx
    =
    \overline S_c,
    \label{eq:mean-preserving}
\end{align}
because \(S_c'\) has zero mean. Thus the limiter enforces the physical bounds at
the points used by the scheme without altering the conservative cell-average
evolution.

After limiting and bound enforcement, the coefficient vector is returned to the
affine constraint space associated with the imposed inflow trace. For \(p>1\),
the production implementation uses the detail-only affine correction introduced
in Eq.~\eqref{eq:detail-affine-correction}. This restores the prescribed trace
without changing the first-cell average. For \(p=1\), no detail subspace is
available, and the code falls back to the full minimum-norm trace correction in
Eq.~\eqref{eq:min-norm-affine-correction}.

Consequently, every Runge--Kutta stage returns a coefficient vector that
satisfies the prescribed inflow trace and preserves the conservative
cell-average update under detail limiting for \(p>1\). The bound-preserving
rescaling enforces the admissible saturation interval at the monitored
quadrature and interface points before the final affine reprojection. In the
constrained inflow cell, the final detail-only reprojection may move some
monitored values outside the admissible interval because the imposed boundary
trace and the preserved first-cell average are enforced simultaneously. In the
flux evaluation, saturation arguments are clipped to the physical interval, and
the trace and mass-balance diagnostics are therefore reported separately in
Sec.~\ref{sec:numerics}.

\subsection{Time integration and CFL condition}
\label{subsec:time-integration}

After the spatial discretization in Eq.~\eqref{eq:mw-weak-residual}, the
cellwise modal/multiwavelet coefficients satisfy the semidiscrete system
\begin{align}
    \frac{d\boldsymbol s}{dt}
    =
    \boldsymbol{\mathcal R}_{\rm tan}(\boldsymbol s),
    \label{eq:semi-discrete-rhs}
\end{align}
where \(\boldsymbol{\mathcal R}_{\rm tan}\) is the conservative residual
projected onto the tangent space of the inflow constraint, as defined in
Eq.~\eqref{eq:tangent-residual-single}. The affine trace constraint is
reimposed after every Runge--Kutta stage to remove roundoff-level drift and to
return the stage value to the admissible coefficient space.

The coefficient system is advanced by the third-order
strong-stability-preserving Runge--Kutta method~\cite{GottliebShuTadmor2001}.
For one step from \(t^n\) to \(t^{n+1}=t^n+\Delta t\), the uncleaned stage
updates are
\begin{align}
    \boldsymbol s^{(1)}
    &=
    \boldsymbol s^n
    +
    \Delta t\,\boldsymbol{\mathcal R}_{\rm tan}(\boldsymbol s^n),
    \label{eq:ssp-rk1}\\
    \boldsymbol s^{(2)}
    &=
    \frac{3}{4}\boldsymbol s^n
    +
    \frac{1}{4}
    \left[
    \boldsymbol s^{(1)}
    +
    \Delta t\,\boldsymbol{\mathcal R}_{\rm tan}(\boldsymbol s^{(1)})
    \right],
    \label{eq:ssp-rk2}\\
    \boldsymbol s^{n+1}
    &=
    \frac{1}{3}\boldsymbol s^n
    +
    \frac{2}{3}
    \left[
    \boldsymbol s^{(2)}
    +
    \Delta t\,\boldsymbol{\mathcal R}_{\rm tan}(\boldsymbol s^{(2)})
    \right].
    \label{eq:ssp-rk3}
\end{align}

After each intermediate stage and after the final stage, the updated state is
processed by the nonlinear cleaning sequence
\begin{align}
    \boldsymbol s
    \longmapsto
    \Pi_{\mathcal A_g}
    \mathcal L_{\rm TVB}
    \mathcal B
    \Pi_{\mathcal A_g}(\boldsymbol s).
    \label{eq:stage-cleaning}
\end{align}
Here \(\Pi_{\mathcal A_g}\) denotes projection onto the affine trace space
\begin{align}
    \mathcal A_g
    =
    \left\{
    \boldsymbol s:\boldsymbol M\boldsymbol s=\boldsymbol g
    \right\},
    \label{eq:affine-trace-space-stage}
\end{align}
\(\mathcal B\) denotes the mean-preserving bound rescaling in
Eq.~\eqref{eq:theta-bound}, and \(\mathcal L_{\rm TVB}\) denotes the
TVB/minmod troubled-cell correction. The first projection enforces the inflow
trace before limiting. The bound-rescaling step then reduces zero-mean detail
content when the reconstructed saturation exceeds the admissible interval at
the monitored quadrature and interface points. The TVB/minmod step further
regularizes troubled cells near shocks. The final projection restores the
affine inflow trace after these nonlinear limiting operations.

For \(p>1\), the final projection is the detail-only correction in
Eq.~\eqref{eq:detail-affine-correction}; for \(p=1\), it reduces to the full
minimum-norm correction in Eq.~\eqref{eq:min-norm-affine-correction}. Thus every
intermediate and final Runge--Kutta state satisfies the prescribed inflow trace.
The mean-preserving bound rescaling controls the monitored saturation values
before the subsequent TVB/minmod and affine reprojection steps. Since the final
detail-only projection enforces the inflow trace by modifying detail
coefficients in the constrained inflow cell, pointwise boundedness in that cell
is monitored numerically rather than guaranteed analytically at every stage.

The time step is chosen from a Courant--Friedrichs--Lewy restriction
~\cite{CourantFriedrichsLewy1967}. For the scalar conservation law
\begin{align}
    \partial_t S+\partial_x\mathcal F(S)=0,
\end{align}
the characteristic speed is
\begin{align}
    a(S)
    =
    \mathcal F'(S)
    =
    \frac{v}{\phi}f'(S).
    \label{eq:characteristic-speed}
\end{align}
We use the global admissible-speed bound
\begin{align}
    a_{\max}
    &=
    \max_{S\in[S_{wc},\,1-S_{or}]}
    |a(S)| \nonumber \\
    &=
    \frac{|v|}{\phi}
    \max_{S\in[S_{wc},\,1-S_{or}]}
    |f'(S)|.
    \label{eq:amax}
\end{align}
For \(p\) local modes per cell, the computations use the explicit time step
\begin{align}
    \Delta t
    =
    C_{\rm CFL}
    \frac{\Delta x_{\min}}
    {(2p+1)a_{\max}},
    \label{eq:cfl}
\end{align}
where \(\Delta x_{\min}=\min_c |I_c|\). The factor \(2p+1\) is the
degree-dependent scaling used here for the explicit high-order discontinuous
transport update, and it becomes more restrictive as the number of local modes
is increased. In the Berea-core simulations reported below, \(p=2\) and
\(C_{\rm CFL}=0.20\). The dependence on \(p\) is examined separately in
Sec.~\ref{subsec:flux-modal-resolution}. The same time-step restriction is used
for residual evaluation, limiting, and affine trace reprojection throughout the
Runge--Kutta stages.

\section{Numerical validation}
\label{sec:numerics}

We validate the conservative affine-constrained multiwavelet coefficient method
on a one-dimensional Berea-core Buckley--Leverett displacement benchmark. The
tests are designed to assess the coefficient-space mechanisms introduced in
Sec.~\ref{sec:mw-conservative}: direct evolution of local modal/multiwavelet
coefficients, conservative interface-flux coupling, affine enforcement of the
inflow trace, detail-only boundary reprojection for \(p>1\), troubled-cell
limiting, and preservation of admissible saturation bounds at the quadrature
and interface points used by the scheme. 

The reference solution is computed independently with \texttt{pywaterflood},
using the same physical parameters, fractional-flow closure, and
pore-volume-injected scaling \cite{Male2024pywaterflood}. The solver used in
this work was implemented directly in Python. The local orthonormal
modal/multiwavelet basis, coefficient-to-node evaluation, conservative weak
residual, numerical interface fluxes, affine boundary lifting, detail limiting,
diagnostics, and explicit time stepping were all implemented in the present
code.

All computations reported here are performed on fixed uniform meshes. The
numerical study therefore focuses on the conservative coefficient-space
mechanisms themselves: affine inflow enforcement, detail-only boundary
reprojection, mean/detail limiting, monotone flux coupling, and mass-balance
preservation. Dynamic mesh refinement, coefficient thresholding, and sparse
adaptive operator representations are left for future extensions.

\subsection{Computational setup}
\label{subsec:berea-setup}

The Berea core has length and diameter
\begin{align}
    L=6.0~\mathrm{in}=0.1524~\mathrm{m},
    \qquad
    D=1.5~\mathrm{in}=0.0381~\mathrm{m},
\end{align}
with cross-sectional area \(A=\pi D^2/4\). The imposed injection rate is
\(q=1.0~\mathrm{mL/min}\), giving the Darcy velocity \(v=q/A\). The porosity and
residual saturations are
\begin{align}
    \phi=0.20,\qquad
    S_{wc}=0.10,\qquad
    S_{or}=0.20 .
\end{align}
Thus the initial and injected saturations are
\begin{align}
    S(x,0)=S_{wc}=0.10,
    \qquad
    S(0,t)=S_{\rm inj}=1-S_{or}=0.80 .
\end{align}
The phase viscosities are
\begin{align}
    \mu_w=1.0\times10^{-3}~\mathrm{Pa\,s},
    \qquad
    \mu_o=4.0\times10^{-3}~\mathrm{Pa\,s}.
\end{align}

For the Berea benchmark, the relative permeabilities are modeled by Corey laws.
The effective water saturation is
\begin{align}
    S_e
    =
    \frac{S-S_{wc}}{1-S_{wc}-S_{or}},
    \qquad
    0\leq S_e\leq1,
    \label{eq:effective-saturation}
\end{align}
and the relative permeabilities are
\begin{align}
    k_{rw}(S) &= k_{rw}^0 S_e^{n_w},\\
    k_{ro}(S) &= k_{ro}^0(1-S_e)^{n_o}.
    \label{eq:corey-relperm}
\end{align}
In the computations reported here,
\begin{align}
    k_{rw}^0=k_{ro}^0=1,
    \qquad
    n_w=n_o=2 .
\end{align}
The phase mobilities are
\begin{align}
    \lambda_w(S)=\frac{k_{rw}(S)}{\mu_w},
    \qquad
    \lambda_o(S)=\frac{k_{ro}(S)}{\mu_o},
\end{align}
and the water fractional flow is
\begin{align}
    f(S)
    =
    \frac{\lambda_w(S)}
    {\lambda_w(S)+\lambda_o(S)} .
    \label{eq:fractional-flow}
\end{align}
During flux evaluation and limiting, \(S_e\) is clipped to \([0,1]\), so that
the computed saturation remains in the physical interval
\begin{align}
    S_{wc}\leq S\leq1-S_{or}.
    \label{eq:saturation-bounds}
\end{align}

The results are reported in terms of pore volumes injected. The pore volume of
the core is \(V_p=\phi A L\), and the dimensionless injected pore volume is
\begin{align}
    \mathrm{PVI}(t)
    =
    \frac{q\,t}{\phi A L}.
    \label{eq:pvi}
\end{align}
Thus one pore volume corresponds to \(t_{\rm PV}=\phi A L/q\). The same
fractional-flow function, physical parameters, initial condition, inflow
saturation, and pore-volume scaling are used in the multiwavelet coefficient
solver and in the \texttt{pywaterflood} reference calculation.

Unless otherwise stated, the simulations use a uniform partition with
\(N_c=256\) cells, \(p=2\) local modes per cell, the Rusanov numerical flux, and
the troubled-cell limiter described in Sec.~\ref{subsec:limiting}. Here \(p=2\)
corresponds to a piecewise-linear local modal representation. This choice is
used as the main production setting because the modal-order sensitivity study
in Sec.~\ref{subsec:flux-modal-resolution} shows that it gives the best
accuracy--cost compromise among the tested orders for the present
shock-dominated benchmark. The time step is chosen from the CFL condition in
Eq.~\eqref{eq:cfl} with \(C_{\rm CFL}=0.20\). The affine trace condition is
reimposed after every Runge--Kutta stage.

\subsection{Breakthrough, profiles, and error measures}
\label{subsec:berea-results}

\begin{figure}[t]
    \centering
    \includegraphics[width=0.48\textwidth]{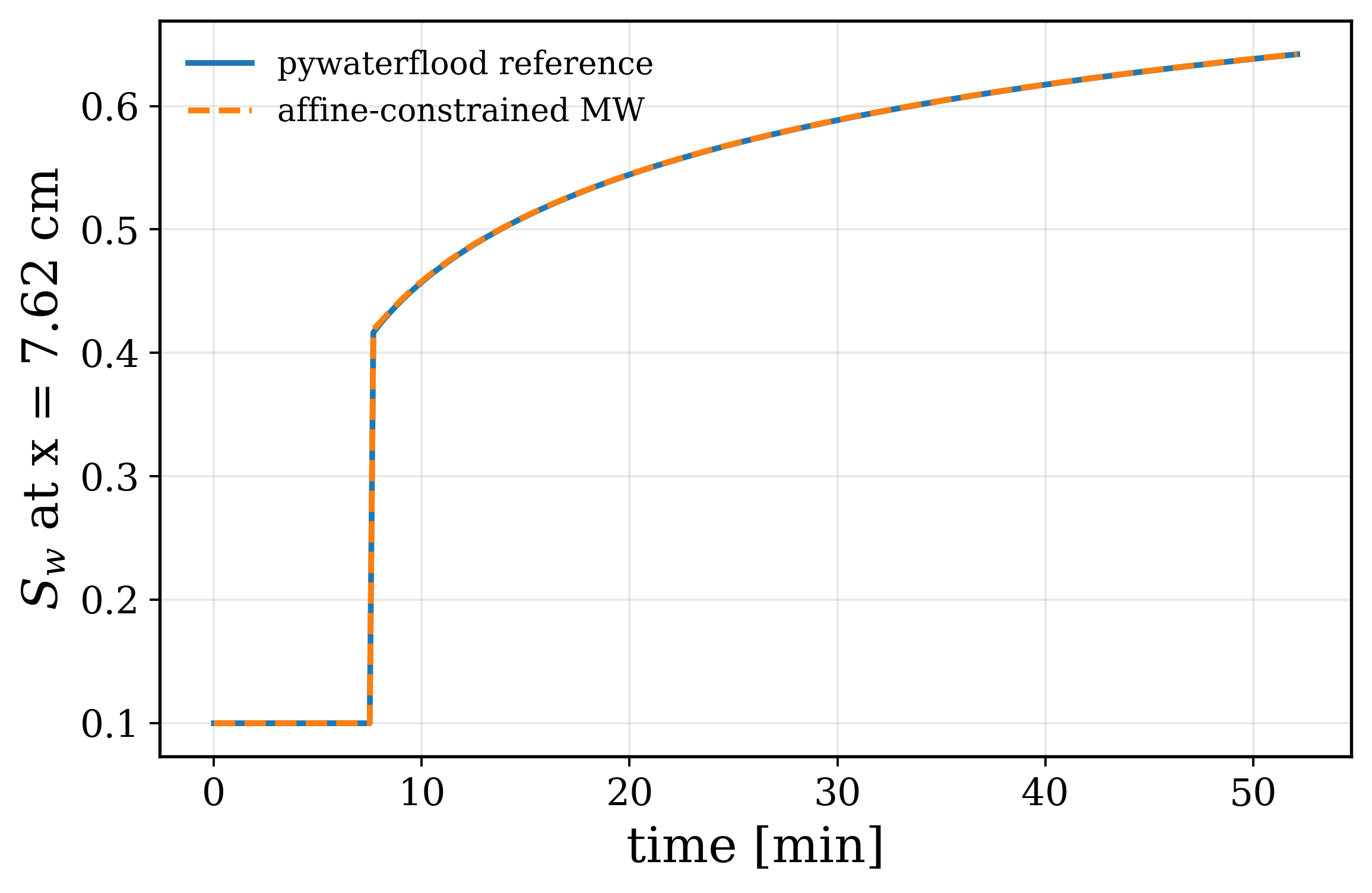}
    \caption{Breakthrough curve for the Berea-core Buckley--Leverett benchmark
    at \(x=L/2=7.62~\mathrm{cm}\). The affine-constrained multiwavelet
    coefficient solution computed with \(N_c=256\), \(p=2\), and Rusanov flux
    is compared with the \texttt{pywaterflood} reference solution using the same
    Corey fractional-flow parameters and pore-volume-injected scaling.}
    \label{fig:probe-breakthrough}
\end{figure}

Figure~\ref{fig:probe-breakthrough} compares the breakthrough behavior at the
midpoint of the core,
\begin{align}
    x_{\rm p}=L/2=7.62~\mathrm{cm}.
\end{align}
Before the arrival of the displacement front, both solutions remain at the
initial connate-water saturation. After breakthrough, the affine-constrained
multiwavelet coefficient solution captures the rapid increase in water
saturation and its subsequent approach toward the injected state. The agreement
of the two curves shows that the conservative interface fluxes propagate the
nonlinear Buckley--Leverett front at the correct speed while the affine
constraint maintains the prescribed inflow saturation.

\begin{figure}[t]
    \centering
    \includegraphics[width=0.48\textwidth]{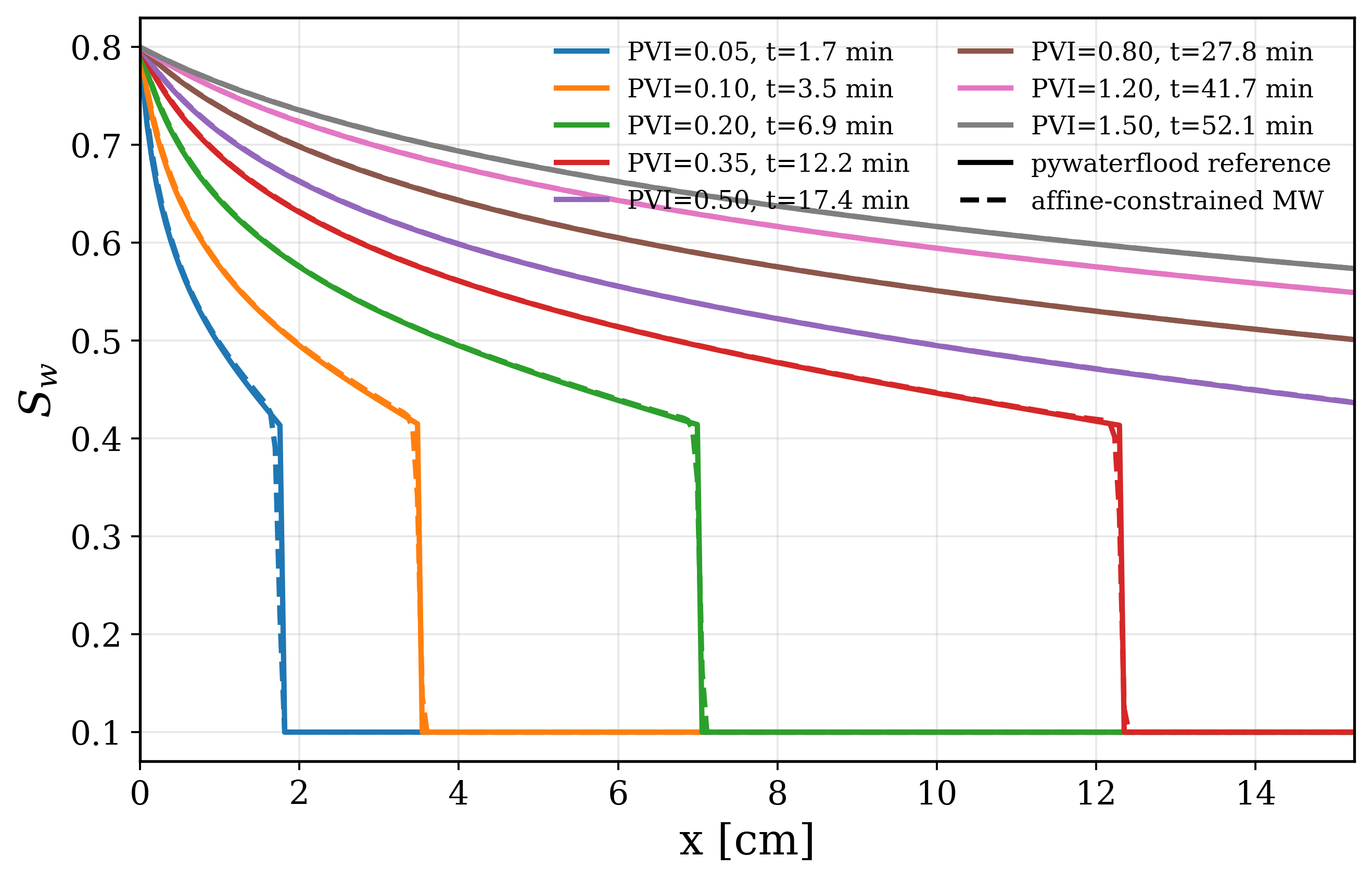}
    \caption{Water-saturation profiles for the Berea-core Buckley--Leverett
    benchmark at selected pore volumes injected. The affine-constrained
    multiwavelet coefficient solution computed with \(N_c=256\), \(p=2\), and
    Rusanov flux is compared with the \texttt{pywaterflood} reference solution.
    The profiles show conservative front propagation, control of saturation bounds
by the detail limiter, and enforcement of the injected-water state at the inflow
boundary.}
    \label{fig:profiles}
\end{figure}

Figure~\ref{fig:profiles} compares water-saturation profiles along the core at
selected pore volumes injected. The affine-constrained multiwavelet coefficient
solution captures the advancing displacement front without visible undershoots
below \(S_{wc}\) or overshoots above \(1-S_{or}\). The profiles remain close to
the reference solution both near the sharp displacement front and in the
smoother post-front saturation region. This comparison also illustrates the
role of the troubled-cell limiter: smooth regions retain the local modal
representation, whereas cells near the shock are locally regularized without
changing their cell averages.

For quantitative comparisons, let \(S_i^{\rm MW}\) denote the
affine-constrained coefficient solution sampled at the cell centers and
\(S_i^{\rm ref}\) the corresponding reference value. We report the
root-mean-square error
\begin{align}
    E_{\rm RMSE}
    =
    \left[
    \frac{1}{N}
    \sum_{i=1}^{N}
    \left(S_i^{\rm MW}-S_i^{\rm ref}\right)^2
    \right]^{1/2},
    \label{eq:rmse-error}
\end{align}
and the maximum pointwise error
\begin{align}
    E_\infty
    =
    \max_i
    \left|S_i^{\rm MW}-S_i^{\rm ref}\right|.
    \label{eq:linf-error}
\end{align}
Because the solution contains a sharp saturation front, \(E_\infty\) is
especially sensitive to small differences in front location, whereas
\(E_{\rm RMSE}\) gives a more global measure of profile agreement.

Table~\ref{tab:berea-errors} reports the profile errors at the same selected
pore volumes used in Fig.~\ref{fig:profiles}. The largest pointwise errors
occur at early pore volumes, when the displacement front is sharp and a small
difference in front position produces a large local discrepancy. In this
regime, two visually close profiles may still give a large \(E_\infty\), because
the error is evaluated pointwise across a nearly discontinuous transition.
After the front has crossed the core, both \(E_{\rm RMSE}\) and \(E_\infty\)
decrease to the \(10^{-4}\) range.

\begin{table}[t]
\caption{Profile errors between the affine-constrained multiwavelet coefficient
solution and the \texttt{pywaterflood} reference solution at selected pore
volumes injected. The calculation uses \(N_c=256\), \(p=2\), and the Rusanov
flux.}
\label{tab:berea-errors}
\begin{ruledtabular}
\begin{tabular}{ccc}
PVI & \(E_{\rm RMSE}\) & \(E_\infty\) \\
\hline
0.05 & \(1.2358\times10^{-2}\) & \(1.94526\times10^{-1}\) \\
0.10 & \(5.2830\times10^{-3}\) & \(7.4713\times10^{-2}\) \\
0.20 & \(5.2850\times10^{-3}\) & \(6.2596\times10^{-2}\) \\
0.35 & \(5.9430\times10^{-3}\) & \(9.0774\times10^{-2}\) \\
0.50 & \(3.8400\times10^{-4}\) & \(7.9200\times10^{-4}\) \\
0.80 & \(2.5100\times10^{-4}\) & \(4.9400\times10^{-4}\) \\
1.20 & \(1.9700\times10^{-4}\) & \(3.2900\times10^{-4}\) \\
1.50 & \(1.7300\times10^{-4}\) & \(2.6300\times10^{-4}\)
\end{tabular}
\end{ruledtabular}
\end{table}

\subsection{Flux comparison, mesh refinement, and modal-order sensitivity}
\label{subsec:flux-modal-resolution}

We next examine the sensitivity of the fixed-grid affine-constrained
multiwavelet coefficient solver to the numerical flux, mesh resolution, and
number of local modes. These tests verify that the agreement shown in
Figs.~\ref{fig:probe-breakthrough} and~\ref{fig:profiles} is not tied to a
single discretization choice. They also quantify the computational tradeoff
between the inexpensive Rusanov flux and the more expensive sampled Godunov
flux.

The parameter \(p\) denotes the number of local modes per cell. Thus \(p=1\)
corresponds to a piecewise-constant representation, \(p=2\) to a
piecewise-linear representation, \(p=3\) to a piecewise-quadratic
representation, and in general the highest local polynomial degree is \(p-1\).
The total number of saturation degrees of freedom is
\begin{align}
    N_{\rm dof}=N_c p .
\end{align}
Because the Buckley--Leverett benchmark contains an entropy shock, the
modal-order study should not be interpreted as a formal smooth-solution
\(p\)-convergence test. In this regime, the global error is controlled by shock
localization, numerical-flux dissipation, and the nonlinear action of the
troubled-cell and bound-preserving limiters.

Table~\ref{tab:flux-comparison} compares the Rusanov and sampled Godunov fluxes
at fixed \(N_c=256\) and \(p=2\). For this benchmark and modal order, the two
monotone fluxes give comparable profile errors, with the Rusanov flux giving
slightly smaller reported global errors and substantially lower wall time. The
sampled Godunov flux remains useful as a monotone comparison flux, but its
additional cost is significant because the local flux extremum is evaluated by
sampling the admissible saturation interval. For this reason, the Rusanov flux
is used as the main production flux in the remaining tests.

\begin{table*}[t]
\caption{Flux comparison at \(N_c=256\) and \(p=2\). The sampled Godunov flux
and the Rusanov flux give comparable errors for this benchmark, while the
Rusanov flux is substantially cheaper in the present implementation.}
\label{tab:flux-comparison}
\begin{ruledtabular}
\begin{tabular}{lccccc}
Flux & \(E_{\rm RMSE}\) & \(E_\infty\) & Trace error & Mass defect & Wall time [s] \\
\hline
Godunov
& \(1.832240\times10^{-4}\)
& \(3.012753\times10^{-4}\)
& \(0.000\times10^{0}\)
& \(1.862\times10^{-10}\)
& \(429.462\) \\
Rusanov
& \(1.730984\times10^{-4}\)
& \(2.629553\times10^{-4}\)
& \(0.000\times10^{0}\)
& \(6.955\times10^{-11}\)
& \(138.092\) \\
\end{tabular}
\end{ruledtabular}
\end{table*}

\begin{figure}[t]
    \centering
    \includegraphics[width=0.48\textwidth]{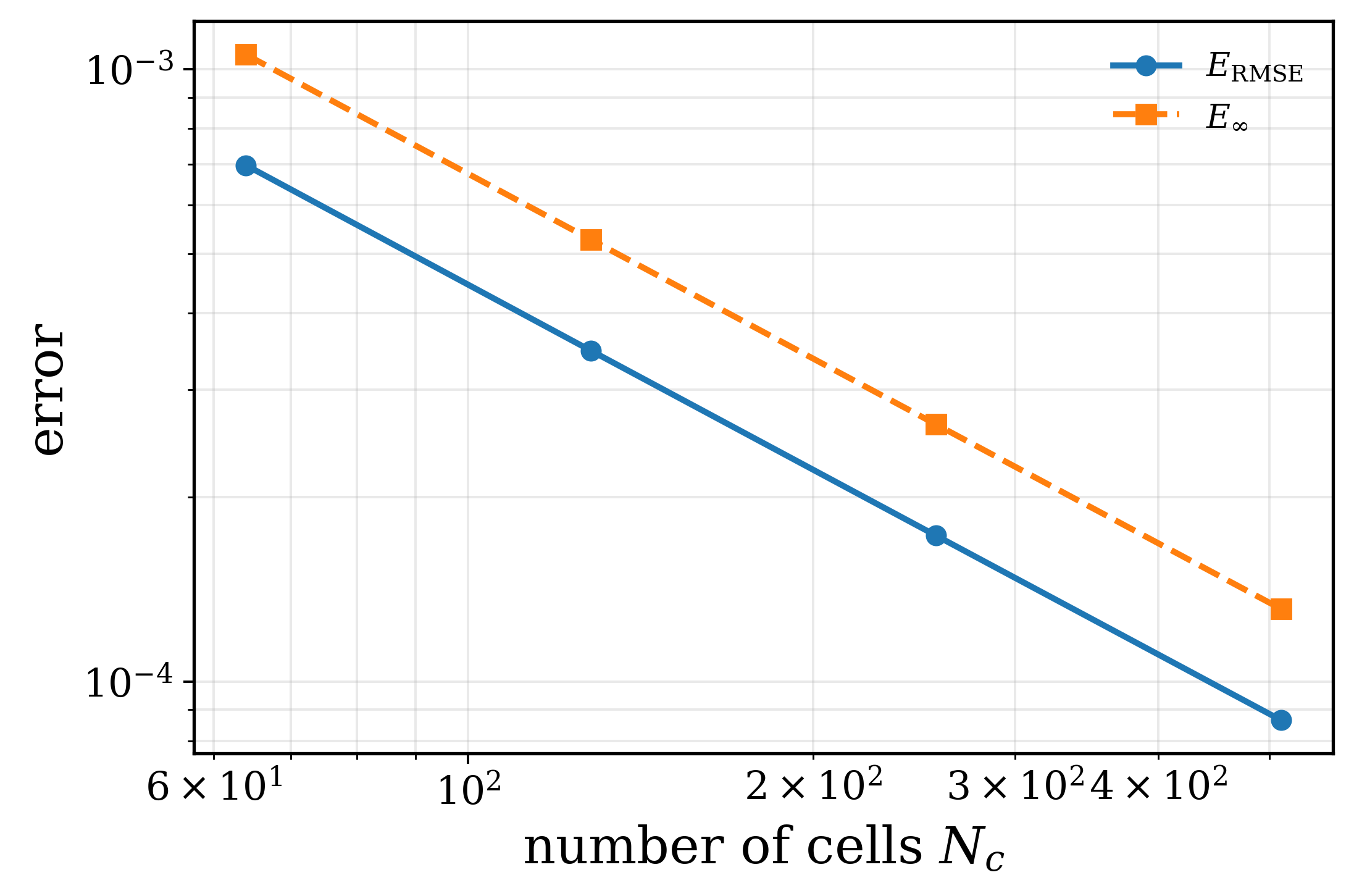}
    \caption{Mesh-refinement study for the affine-constrained multiwavelet
    coefficient solver using the Rusanov flux and \(p=2\). Both the
    root-mean-square error and maximum pointwise error decrease under
    refinement. The maximum error is dominated by the narrow shock region and is
    therefore more sensitive to small front-location differences.}
    \label{fig:resolution-study}
\end{figure}

Figure~\ref{fig:resolution-study} reports the mesh-refinement behavior at fixed
\(p=2\) using the Rusanov flux. Both the root-mean-square error and the maximum
pointwise error decrease under refinement. The decrease of \(E_{\rm RMSE}\)
shows systematic improvement of the global saturation profile, while
\(E_\infty\) is more sensitive to small differences in front location because
the reference profile contains a narrow shock transition.

\begin{figure}[t]
    \centering
    \includegraphics[width=0.48\textwidth]{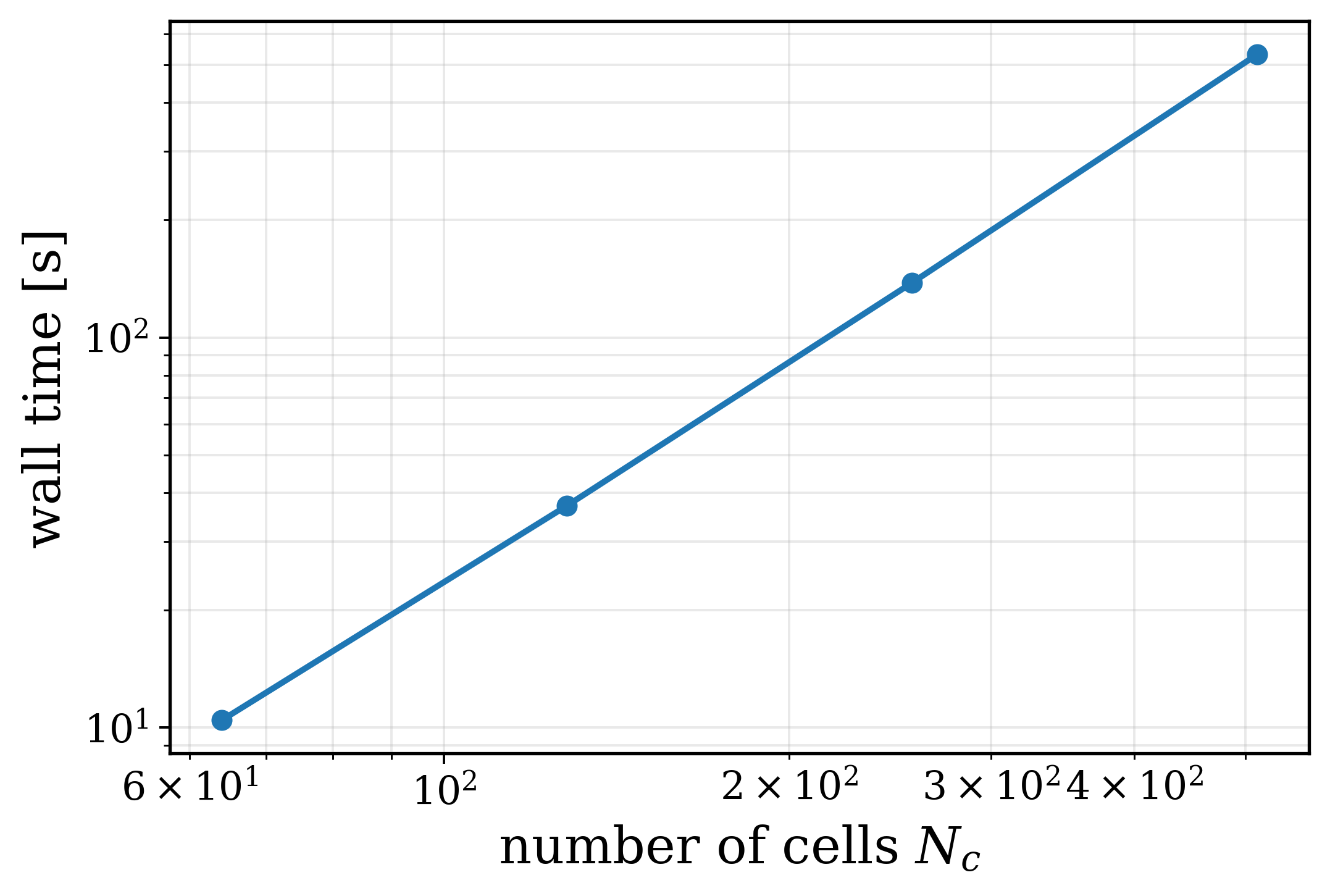}
    \caption{Runtime scaling for the affine-constrained multiwavelet
    coefficient solver using the Rusanov flux and \(p=2\). For fixed final
    physical time, the wall-clock time increases superlinearly with \(N_c\),
    because refinement increases both the number of degrees of freedom and the
    number of explicit time steps through the CFL restriction.}
    \label{fig:resolution-runtime}
\end{figure}

The associated wall-clock time is shown in
Fig.~\ref{fig:resolution-runtime}. For a fixed final physical time, the runtime
increases superlinearly with \(N_c\), because mesh refinement increases both the
number of degrees of freedom and the number of explicit time steps through the
CFL condition. The timings are reported for the present Python implementation
and are intended to quantify relative cost across the tested discretizations,
not optimized parallel scalability.

Table~\ref{tab:modal-order-sensitivity} reports the modal-order sensitivity at
fixed \(N_c=256\) using the Rusanov flux. The dependence on \(p\) is
nonmonotone, as expected for a shock-capturing calculation with nonlinear
limiting. In smooth regions, increasing \(p\) enriches the local representation.
Near the Buckley--Leverett front, however, the error is dominated by front
localization and by the interaction between modal detail coefficients and the
troubled-cell limiter. In the present benchmark, \(p=2\) gives the most
favorable accuracy--cost compromise among the tested modal orders; \(p=4\)
gives comparable accuracy at substantially larger runtime, while \(p=3\) is
more sensitive to limiter--front interaction.

\begin{table*}[t]
\caption{Modal-order sensitivity at fixed \(N_c=256\) using the Rusanov flux.
Here \(p\) is the number of local modes per cell and the polynomial degree is
\(p-1\). Because the solution contains a shock, monotone improvement with
increasing \(p\) is not expected.}
\label{tab:modal-order-sensitivity}
\begin{ruledtabular}
\begin{tabular}{ccccccc}
\(p\) & Degree & \(N_{\rm dof}\) & \(N_{\rm steps}\) &
\(E_{\rm RMSE}\) & \(E_\infty\) & Wall time [s] \\
\hline
1 & 0 & 256  & 19190 & \(9.351175\times10^{-3}\) & \(1.405022\times10^{-2}\) & 4.887 \\
2 & 1 & 512  & 31983 & \(1.730984\times10^{-4}\) & \(2.629553\times10^{-4}\) & 138.092 \\
3 & 2 & 768  & 44775 & \(1.787295\times10^{-3}\) & \(1.687640\times10^{-2}\) & 195.469 \\
4 & 3 & 1024 & 57568 & \(2.395573\times10^{-4}\) & \(1.903429\times10^{-3}\) & 251.171 \\
\end{tabular}
\end{ruledtabular}
\end{table*}

Finally, we monitor the diagnostics most directly connected with the proposed
coefficient-space construction: the imposed inflow trace and the discrete
global mass balance. The affine trace constraint is designed to keep the left
boundary value fixed at the injected saturation, while the conservative weak
residual gives the discrete global mass balance up to time-integration and
quadrature error. These diagnostics are essential because the method combines
three nonlinear operations at every Runge--Kutta stage: flux evaluation, detail
limiting, and affine trace reprojection.

\begin{figure}[t]
    \centering
    \includegraphics[width=0.48\textwidth]{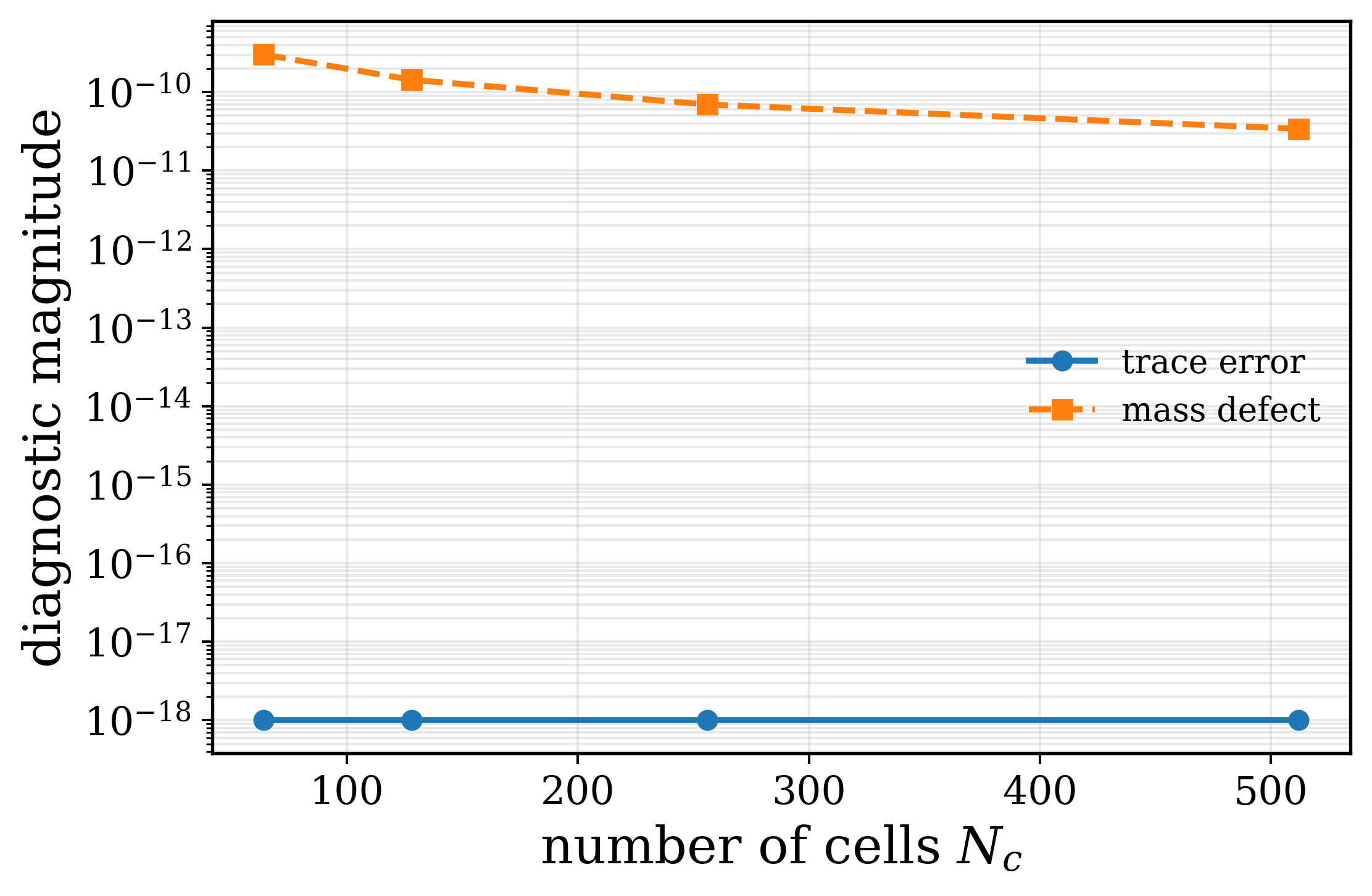}
    \caption{Boundary-trace and mass-balance diagnostics for the
    affine-constrained multiwavelet coefficient solver using the Rusanov flux
    and \(p=2\). The imposed inflow trace is preserved at roundoff level, while
the accumulated global mass-balance defect remains small over the refinement
study.}
    \label{fig:constraint-diagnostics}
\end{figure}

The trace error is measured as
\begin{align}
    E_{\rm tr}(t)
    =
    \left|S_h(0,t)-S_{\rm inj}\right|,
    \label{eq:trace-error}
\end{align}
and the accumulated mass-balance defect is computed from the difference between
the change in total saturation and the time-integrated net boundary flux,
\begin{align}
    E_M(t)
    & =
    \Bigg|
    \int_\Omega S_h(x,t)\,dx
    -
    \int_\Omega S_h(x,0)\,dx  \nonumber \\
    &+
    \int_0^t
    \Big[
    \widehat{\mathcal F}_{N_c+1/2}(\tau)
    -
    \widehat{\mathcal F}_{1/2}(\tau)
    \Big]d\tau
    \Bigg|.
    \label{eq:mass-defect}
\end{align}
Figure~\ref{fig:constraint-diagnostics} shows that the affine inflow constraint
is preserved to roundoff accuracy and that the accumulated mass-balance defect
remains small over the refinement study. These diagnostics confirm that the
coefficient-space boundary treatment is compatible with the conservative weak
residual, even though the stage update combines nonlinear flux evaluation,
detail limiting, and affine trace reprojection.

\section{Conclusions}
\label{sec:conclusions}

We have developed a fixed-grid conservative affine-constrained
modal/multiwavelet coefficient method for the one-dimensional
Buckley--Leverett saturation equation. The saturation is evolved directly in a
local orthonormal coefficient space, and the nonlinear fractional-flow front is
advanced through a conservative weak formulation with monotone numerical
interface fluxes. The method therefore treats the coefficient representation as
the primary discretization space, rather than as a post-processing
reconstruction, compression device, or error indicator.

The main contribution is the coefficient-space treatment of the hyperbolic
inflow condition. The prescribed inflow trace is imposed as a linear constraint
on the coefficient vector and enforced by affine lifting. For \(p>1\), the
stage reprojection is applied in the detail subspace of the inflow cell, so that
the trace is restored without changing the mean coefficient and hence without
altering the conservative cell-average update. Shock-induced oscillations are
controlled by a troubled-cell limiter acting on zero-mean detail coefficients,
while saturation bounds are controlled by a mean-preserving rescaling at the
quadrature and interface points used by the scheme. This construction keeps the
boundary constraint, the limiter, and the conservative update in the same
coefficient space.

The Berea-core benchmark validates the complete fixed-grid formulation. Using
the same Corey fractional-flow closure, physical parameters, and
pore-volume-injected scaling as the independent \texttt{pywaterflood} reference
calculation, the affine-constrained coefficient solver reproduces the
breakthrough curve and saturation profiles. The imposed inflow trace is
preserved to roundoff accuracy, the accumulated mass-balance defect remains
small, and mesh refinement at fixed \(p=2\) reduces the profile errors. The flux
and modal-order studies show that the Rusanov flux with \(p=2\) provides the
most favorable accuracy--cost compromise among the tested settings for this
shock-dominated benchmark.

The present implementation is intentionally fixed-grid. It does not use dynamic
mesh refinement, coefficient thresholding, or sparse adaptive operator
representations. Its purpose is to establish the conservative
coefficient-space ingredients needed before introducing adaptivity: affine
inflow enforcement, detail-only trace correction, mean/detail limiting,
saturation-bound control, and conservative flux coupling. These validated
operations provide a basis for future extensions to hierarchical adaptive
meshes, local \(hp\)-adaptivity, heterogeneous cores, gravity and capillary
corrections, and coupled pressure--transport simulations.

\section*{Data availability}
The data supporting the findings of this study are available from the corresponding author upon reasonable request. The code used in this work is publicly available at \url{https://github.com/Christian48596/mw_affine_buckley_leverett}. The repository contains the Buckley--Leverett multiwavelet solver, the default Berea benchmark settings, and documentation for modifying the rock and fluid parameters. The results reported in this study can be reproduced from the code and input settings provided in that repository.

\begin{acknowledgments}
Ch.T. acknowledges support from the Deanship of Research (DOR) at King Fahd University of Petroleum \& Minerals (KFUPM) through project No.~EC251017.
E.D. acknowledges support from the Research Council of Norway through its
Centres of Excellence scheme, Hylleraas Centre, project No.~262695. The authors
thank Prof. Dr. Antoine Levitt, Laboratoire de Mathématiques d'Orsay at the Université
Paris-Saclay, and Dr. Magnar Bj\o{}rgve. UiT The Arctic University of Norway, for useful discussions.
\end{acknowledgments}

\bibliography{main}

\end{document}